\documentclass[conference]{IEEEtran}

\usepackage[cmex10]{amsmath}
\usepackage{amssymb,amsthm}
\usepackage{graphicx}

\usepackage{amsmath,amsfonts}
\usepackage{subfigure}
\usepackage[small]{caption}
\usepackage{mathrsfs}
\usepackage{booktabs}
\usepackage[all,2cell,ps]{xy}
\newcommand{\god}{G\"odel}

\newcommand{\R}{{\mathbb{R}}}

\newcommand{\nmm}{NM\textsuperscript{-}}
\newcommand{\NMA}{{\mathbf{NM}}} 

\newcommand{\Form}{\textsc{Form}}

\renewcommand{\d}{\ar@{-}[d]}
\renewcommand{\u}{\ar@{-}[u]}
\newcommand{\dr}{\ar@{-}[dr]}
\newcommand{\ur}{\ar@{-}[ur]}
\newcommand{\dl}{\ar@{-}[dl]}
\newcommand{\ul}{\ar@{-}[ul]}
\newcommand{\dll}{\ar@{-}[dll]}
\newcommand{\dtl}{\ar@{-}[dlll]}
\newcommand{\dql}{\ar@{-}[dllll]}
\newcommand{\dcl}{\ar@{-}[dlllll]}
\newcommand{\dsl}{\ar@{-}[dllllll]}
\newcommand{\dstl}{\ar@{-}[dlllllll]}
\newcommand{\drr}{\ar@{-}[drr]}
\newcommand{\dtr}{\ar@{-}[drrr]}
\newcommand{\dqr}{\ar@{-}[drrrr]}
\newcommand{\dcr}{\ar@{-}[drrrrr]}
\newcommand{\dsr}{\ar@{-}[drrrrrr]}
\newcommand{\ull}{\ar@{-}[ull]}
\newcommand{\urr}{\ar@{-}[urr]}

\newtheorem{theorem}{Theorem}[section]
\newtheorem{lemma}[theorem]{Lemma}

\newtheorem{proposition}[theorem]{Proposition}

\newtheorem{definition}[theorem]{Definition}
\theoremstyle{definition}
\newtheorem*{remark}{Remark}

\newtheorem{example}{Example}

\begin{document}
%
\title{Valuations in Nilpotent Minimum Logic}

\author{\IEEEauthorblockN{Pietro Codara}
\IEEEauthorblockA{Dipartimento di Informatica\\
Universit\`{a} degli Studi di Milano\\
Milan, Italy\\
e-mail: codara@di.unimi.it}
\and
\IEEEauthorblockN{Diego Valota}
\IEEEauthorblockA{
Artificial Intelligence Research Institute (IIIA)\\ 
CSIC\\
Campus UAB, 08193 Bellaterra, Spain\\
e-mail: diego@iiia.csic.es
}}
\maketitle

\begin{abstract}
The Euler characteristic can be defined as a special kind of valuation on finite distributive
lattices. This work begins with some brief consideration on the r\^{o}le of the Euler
characteristic on NM algebras, the algebraic counterpart of Nilpotent Minimum logic.
Then, we introduce a new valuation, a modified version
of the Euler characteristic we call \emph{idempotent Euler characteristic}.
We show that the new valuation encodes  information about the formul\ae{} in NM
propositional logic.
\end{abstract}

{\keywords NM logic; NM algebra; \nmm{} logic; valuation; Euler characteristic}

\section{Introduction}\label{sec:intro}

Let $L$ be a
distributive
lattice. A function $\nu \colon L \to \R$ is a
\emph{valuation} if it satisfies
\begin{equation}\label{eq:valuation}
 \nu(x)+\nu(y)=\nu(x\vee y)+\nu(x\wedge y)
\end{equation}
for all $x,y,z \in L$.  Recall that an element $x \in L$ is
\emph{join-irreducible} if it is not the bottom element of $L$, and $x = y
\vee z$ implies $x=y$ or $x=z$ for all $y,z \in L$. When $L$ is finite, it turns
out \cite[Corollary 2]{rota} that any valuation $\nu$
is uniquely determined by its values on the join-irreducible elements of
$L$, along with its value at the bottom element $\bot$ of $L$.

A special kind of valuation, introduced by V. Klee and G.-C.\ Rota, is the Euler characteristic,
defined as follows.

\begin{definition}[{\cite[p.\ 120]{klee}, \cite[p.\ 36]{rota}}] The Euler
characteristic of a finite
distributive lattice $L$ is the unique  valuation $\chi \colon L \to
\R$ such
that $\chi(x)=1$ for any join-irreducible element $x \in L$, and $\chi(\bot)=0$.
\end{definition}

In \cite{cdm_ismvl,cdm_jmvlsc}, the authors investigate the notion of Euler characteristic in a
particular case of finite distributive lattice: G\"{o}del algebras, the algebraic counterpart
of the many-valued logic known as G\"{o}del logic\footnote{For background on G\"{o}del logic
see, e.g., \cite{hajek}. The characterization of G\"{o}del algebra used in the cited papers
is provided in \cite{adm_jlc,cdm_ijar,dm_apal}.}. Specifically, they consider the
Lindenbaum algebra of G\"{o}del logic over a finite set of variables and then they investigate
the values assigned by the Euler characteristic to each equivalence class of formul\ae. It turns
out that the Euler characteristic encode logical information about the formul\ae, but such information
is classical, i.e. coincide with the analogous notion defined in classical propositional logic;
namely, the Euler characteristic of a formula is the number of Boolean assignments which makes the formula
true. Further, the authors generalize the notion of Euler characteristic to a family of new valuations,
the many-valued versions of the Euler characteristic. The latter valuations are shown to be able
to separate many-valued tautologies from non-tautologies.

In this paper we approach the same problem on a different many-valued logic, the Nilpotent Minimum logic NM.
We will briefly investigate the logical meaning of the Euler characteristic on NM algebras, the algebraic
counterpart of NM logic, showing that such valuation, as is, can not carry information about assignments making
a formula classically true. In order to obtain such a result we will introduce a new valuation, a modified version
of the Euler characteristic we call \emph{idempotent Euler characteristic}, and prove that such valuation indeed
is capable of capturing the desired information.

The NM logic is briefly presented in the next section. Section \ref{s:main} contains
our main results. In Section \ref{s:NM-} we spend a few word to describe a particular schematic
extension of NM logic, known as the logic \nmm.
We easily obtain, as a corollary of our main result, that the \emph{idempotent Euler characteristic}
on \nmm{} algebras plays exactly the same r\^{o}le as the Euler characteristic on \god{}
algebras. We conclude our work with some consideration on possible further results.

\section{The logic of the Nilpotent Minimum}\label{sec:NM}

A \emph{triangular norm} (also called \emph{t-norm}; see \cite{KMP00}) is a binary, commutative, associative and
monotonically non-decreasing operation on $[0,1]^2$ that has $1$ as unit element.
The \emph{Nilpotent Minimum} t-norm is a first example of a left-continuous but
not continuous  t-norm.
It has been introduced by Fodor \cite{F95}, and it is defined as
\begin{align}
x \odot y  &= \begin{cases}
		min\{x,y\} & \text{if } x + y > 1,\\
		0 & \text{otherwise.}
		\end{cases}  \label{eq:nmtnorm}
\end{align}
for every $x,y\in[0,1]$.

Hence, the \emph{Nilpotent Minimum propositional logic} (NM for short) lies in
the hierarchy of extensions of the \emph{Monoidal T-norm based Logic} (MTL),
introduced in \cite{EG01} by Esteva and Godo.
The propositional language of MTL is built over the binary connectives $\odot,\wedge,\to$
and the constant $\bot$.
Usually derived connectives are $x\leftrightarrow y=(x\to y)\odot(y\to x)$,
$x \vee y = ((x\to y)\to y)\wedge((y\to x)\to x)$,
the negation $\neg x = x\to\bot$,
and the constant $\top = \neg\bot$. We let $\varphi^2=\varphi\odot\varphi$.

The WNM logic is obtained from MTL by adding the axiom:
\begin{align*}
\neg(x \odot y) \vee ((x \wedge y) \to (x \odot y)),   \tag{WNM}\label{ax:wnm}  \\
\end{align*}
while NM logic is given by WNM plus involutivity axiom:
\begin{align*}
\neg\neg x \to x. \tag{INV}\label{ax:nm}
\end{align*}
The aforementioned G\"{o}del logic
can be obtained by adding the idempotency axiom to MTL logic.
If we add the axiom
$\neg(\neg x^2)^2 \leftrightarrow (\neg(\neg x)^2)^2$
to NM, we obtain its negation fixpoint-free version, called NM$^-$ \cite{G03}.

The following form of \emph{local} deduction theorem holds in NM logic \cite{ABM07},
\begin{equation}\label{eq:deduction}
\varphi\vdash_{NM}\psi  \text{ if and only if } \vdash_{NM}\varphi^2\to\psi\text{.}
\end{equation}
Hence, we say that NM logic \emph{proves} $\psi$ from $\varphi$, in symbols $\varphi\vdash_{NM}\psi$,
when $\varphi^2\to\psi$ is a theorem of NM logic.

The algebraic semantic of MTL is given by the variety of \emph{MTL algebras} \cite{EG01}.
As G\"{o}del algebras are exactly the prelinear Heyting algebras,
NM algebras are the prelinear Nelson algebras \cite{BC10}.
Hence, NM logic is to \emph{Nelson logic} (constructive logic with strong negation)
as G\"{o}del logic is to Intuitionistic logic.

The algebraic variety of NM algebras corresponding to NM logic has a nice property,
that it is \emph{locally finite} \cite{NEG08}.
This means that finitely generated free algebras are finite.
Hence, a combinatorial treatment of free $n$-generated algebras is feasible.
Indeed, 
a characterization of free $n$-generated NM algebras based on partially ordered sets
(posets for short) has been given in \cite{AG08}.

In the next section we introduce some algebraic and combinatorial notion that will be
useful throughout the paper.

\subsection{NM algebras}\label{sec:background}

Abusing notation, in the following we identify logical connectives with their
algebraic interpretations.
An algebra $\mathbf{A}=\langle A,\wedge,\vee,\odot,\to,\bot,\top\rangle$ of type $(2,2,2,2,0,0)$
is a WNM algebra if and only if $(A,\wedge,\vee,\bot,\top)$ is a bounded lattice, with top $\top$ and bottom $\bot$,
$\langle A,\odot,\top\rangle$ is a commutative monoid, and it satisfies the \emph{residuation} equation,
$x \odot y \leq z$ if and only if $x \leq y \to z$, the \emph{prelinearity} equation $(x \to y ) \vee (y \to x)=\top$,
the \emph{weak nilpotent minimum} equation $\neg(x \odot y) \vee ((x \wedge y) \to (x \odot y))=\top$.
Therefore, WNM algebras are a class of involutive residuated lattices.
When the lattice order is total, $\mathbf{A}$ is called a \emph{chain}.
A WNM algebra that satisfies the \emph{involutivity} equation $(x\to\bot)\to\bot = x$ is called \emph{NM algebra},
while a \emph{G{\"o}del} algebra is an WNM algebra that satisfies \emph{idempotency}, that is $x \odot x = x$.
Negation $\neg x$ is usually defined by $x \to \bot$.
An NM algebra satisfying  $\neg(\neg x^2)^2 \leftrightarrow (\neg(\neg x)^2)^2=\top$ is called
a \emph{NM$^{-}$ algebra}.
Given an element $x$ of a NM algebra $\mathbf{A}$, we say that $x$ is \emph{negative} when $x<\neg x$,
$x$ is \emph{positive} when $x>\neg x$. We call $x$ a \emph{negation fixpoint} when $x=\neg x$.
Note that if $\mathbf{A}$ has a negation fixpoint, then it is unique.

The variety $\mathbb{NM}$ of NM algebras is generated by the \emph{standard} NM algebra
$\mathbf{[0,1]}=\langle [0,1],\wedge^{[0,1]},\vee^{[0,1]},\odot^{[0,1]},\to^{[0,1]},0,1\rangle$
where $\odot^{[0,1]}$ is the NM t-norm \eqref{eq:nmtnorm},
$x\wedge^{[0,1]} y = min\{x,y\}$, $x\vee^{[0,1]} y = max\{x,y\}$ and
\begin{align}
x \to^{[0,1]} y &=  \begin{cases}
		1 & \text{if } x \leq y \\
		max\{\neg x,y\} & \text{otherwise.}
	\end{cases}\label{eq:nmres}
\end{align}
for every $x,y\in[0,1]$.

By the subdirect representation theorem \cite{BS81} and the fact that subdirectly irreducible
MTL algebras are chains \cite{EG01}, every NM algebra $\mathbf{A}$ is isomorphic to a subdirect product
of a family $(C_i)_{i\in I}$ of NM chains, for some index set $I$.
When $\mathbf{A}$ is finite and not trivial, then the family $(C_i)_{i\in I}$ of non trivial chains
is essentially unique up to reordering of the finite index set $I$.
Hence, there exist $\pi_i:\mathbf{A}\to C_i$ such that
$\pi_i(a)=a_i$ for every $a\in\mathbf{A}$. We call $a_i$ the \emph{$i^{th}$-projection} of $a$.
Then, we can display every element $a$ in $\mathbf{A}$ by means of its projections $(a_i)_{i\in I}$.

Since every finite NM chain $C=\langle C,\odot,\to, \vee,\wedge,\bot,\top\rangle$ is
a subalgebra of $\mathbf{[0,1]}$, then  by \eqref{eq:nmtnorm} and \eqref{eq:nmres}
and the fact that $\neg^{[0,1]} x := x \to^{[0,1]} 0$, we have
\begin{align}
x \odot y &= \begin{cases}
		min(x,y) & x > \neg y; \\
		\bot 	& x \leq \neg y.
\end{cases}\label{eq:nmch} \\
x \to y &= \begin{cases}
	\top 	& x \leq y; \\
	max(\neg x,y) 	& x>y.
	\end{cases}\label{eq:nmresch}
\end{align}
for all $x,y\in C$.

Note that, given a NM chain $C$, every $x\in C$ is either positive, negative or a negation fixpoint.

Denote by $\Form_n$ the set of all well-formed formul\ae{} of NM logic whose
propositional variables are contained in $\{x_1,...,x_n\}$.
Let $\mathbf{A}$ be a NM algebra, with $a_1,...,a_n\in A$, and let $\varphi\in \Form_n$.
By $\varphi^A(a_1,...,a_n)$ we denote the element of $A$ obtained by the evaluation of $\varphi$ in $A$ interpreting
every $x_i$ with the corresponding $a_i$, in particular $x_i^A=a_i$.
With this notation a formula $\varphi$ is a \emph{tautology} of NM logic
if and only if for every algebra $A\in\mathbb{NM}$
and for every $a_1,...,a_n\in A$, $\varphi^A(a_1,...,a_n)= \top^A$.
Moreover, given two logical formul\ae{} $\varphi$ and $\psi$, we say that they are \emph{logically equivalent}
if and only if $(\varphi\leftrightarrow\psi)^A = \top^A$, for every $A\in\mathbb{NM}$.
In symbols, $\varphi\equiv\psi$. Note that $\equiv$ is an equivalence relation.
The algebra whose elements are the equivalence classes of formul\ae{} of NM logic with respect to
$\equiv$ is called the \emph{Lindenbaum Algebra} of NM and its elements are denoted
$[\varphi]_{\equiv}$.
The free $n$-generated algebra $\NMA_n$ in $\mathbb{NM}$ is the Lindenbaum algebra
of the logical formul\ae{} over the first $n$ variables.
Since $\mathbf{[0,1]}$ is generic for $\mathbb{NM}$, then $\NMA_n$ is isomorphic to the subalgebra
of $[0,1]^{[0,1]^n}$ generated by the projection functions $(a_1,\dots,a_n)\mapsto a_i$.
It follows that there exists a map from equivalence classes of formul\ae{} $[\varphi]_{\equiv}$
to real-valued functions $f:[0,1]^n\to[0,1]$.

Given a finite poset $F$ and $S\subseteq F$, the \emph{lower set} of $S$ is
$\downarrow S=\{x\in F\mid x\leq y \text{ for some } y\in S\}$, and
the \emph{upper set} of $S$ is
$\uparrow S=\{x\in F\mid x\geq y \text{ for some } y\in S\}$.
A \emph{forest} is a finite poset such that for every $x\in F$ the lower set $\downarrow\{x\}$ is a chain.
A forest with a bottom element is called a \emph{tree}, and its bottom element
is called \emph{root}.

Let $\mathbf{A}$ be a finite NM algebra.
A nonempty subset $S$ of $A$ is called a \emph{filter} of $\mathbf{A}$ when
$S$ is an upper set, and
for all $x,y \in S$ then $x \odot y \in S$ .
Since $S$ is finite, then it has a minimum element $\bigwedge_{x \in S}x$
(that is, $S$ is principal).
We call \emph{generator} of $S$ the minimum element of the filter $S$.
A filter $S$ of $A$ is \emph{prime} if $S \neq A$ and for all $x,y \in A$, $x \vee y\in S$
implies $x\in S$ or $y\in S$.
Note that, for every prime filter $S$ of $\mathbf{A}$,
its generator is an idempotent join irreducible element of $\mathbf{A}$.
We consider the reverse inclusion as a partial order between prime filters,
that is $S\leq S'$ if and only if $S'\subseteq S$, for every couple of filters $S$ and $S'$.

\begin{proposition}[\cite{ABM07}]\label{prop:specfree}
The set of prime filters of $\NMA_n$ ordered by reverse inclusion is a forest.
\end{proposition}

As a direct consequence of Proposition \ref{prop:specfree},
when $S$ is generated by a minimal idempotent join irreducible elements of $\NMA_n$,
then $S$ is the root of a tree in the forest of prime filters of $\NMA_n$. In such case, following
the classical terminology, we say that $S$ is maximal (with respect to the inclusion among filters).

We conclude the Section with a simple Lemma
\footnote{We thank the anonymous  referee for pointing out that Lemma \ref{lem:maxquotient} can be generalized to any NM-algebra, and not just to
finite ones. This follows from the fact that the quotient by a maximal
filter is a simple algebra and that up to isomorphism the only simple
NM-algebras are $\mathbf{2}$ and $\mathbf{3}$.}
%
that will be useful in the following.

\begin{lemma}\label{lem:maxquotient}
Let $\mathbf{2}$ and $\mathbf{3}$ be the two-elements and the three-elements NM chains, respectively.
Then, given a finite NM algebra $\mathbf{A}$ and a maximal prime filter $\mathbf{p}$,
the quotient $\mathbf{A}/\mathbf{p}$ is either isomorphic to $\mathbf{2}$, or isomorphic to $\mathbf{3}$.
\end{lemma}
\begin{proof}
Let $(C_i)_{i\in I}$ be the subdirect representation of $\mathbf{A}$,
and let $p\in \mathbf{A}$ be the join irreducible element that generates $\mathbf{p}$.
Note that since $\mathbf{p}$ is maximal and prime, then $p$ is minimal and idempotent.

Since $p$ is join irreducible then there exists only one $j\in I$ such that $p_j\not=\bot_j$.
Moreover, $p_j>\neg p_j$, for else $p_j\odot p_j=\bot_j$, in contradiction with the idempotency of $p$.
Finally, since $p$ is minimal, $p_j$ is the least positive element in $C_j$. Moreover,
if $C_j$ does not have a negation fixpoint $f$, $\neg p_j$ is the greatest negative element in $C_j$, otherwise
$f$ covers $\neg p_i$.

Denote with $\sim_{\mathbf{p}}$ the congruence associated to $\mathbf{p}$.
By the above discussion, if $C_j$ does not have a negation fixpoint then $\NMA_n/\mathbf{p}$ is isomorphic to
the two element NM chain $[p]_{\sim_{\mathbf{p}}}>[\neg p]_{\sim_{\mathbf{p}}}$.
Otherwise, if $C_j$ has a negation fixpoint $f$ then $\mathbf{A}/\mathbf{p}$ is isomorphic to
the three element NM chain $[p]_{\sim_{\mathbf{p}}}>[f]_{\sim_{\mathbf{p}}}>[\neg p]_{\sim_{\mathbf{p}}}$.
\end{proof}

\section{Valuations in NM logic}\label{s:main}


\smallskip Since $\NMA_n$ is a finite distributive lattice whose elements
are formul\ae\ in $n$ variables, up to logical equivalence, we can extend the scope
of valuations to formul\ae, as follows.

\begin{definition}\label{d:val_on_NM algebra}Let $\nu:\NMA_n \to \R$ be a valuation
on the finite distributive lattice $\NMA_n$. The \emph{valuation} $\nu(\varphi)$
\emph{of a formula} $\varphi \in \Form_n$ is the number $\nu([\varphi]_\equiv)$.
\end{definition}


As mentioned in the introduction, one of the goals of \cite{cdm_jmvlsc} is the interpretation
of the logical meaning of the Euler characteristic on G\"{o}del algebras.
In that specific case, it turns out that the Euler characteristic of a formula $\varphi$
coincide with the number of Boolean assignments satisfying $\varphi$.

Turning now to the case of NM-algebras, we can hope that the Euler
characteristic $\chi(\varphi)$ of a formula $\varphi$ encodes information
about assignments making $\varphi$ true.
At least, this should work for the join irreducible elements
of $\NMA_n$. But, unfortunately, this is not the case.
Indeed, take, for instance, the formula
$$\alpha = (X \leftrightarrow \neg X)^2 \wedge X\,.$$

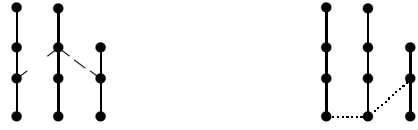
\begin{figure}[h!]
\begin{center}\leavevmode
\xymatrix@C=0.9pc@R=1pc{
*={\bullet}\d			&*={\bullet}\d			&	 \\
*={\bullet}\d			&*={\bullet}\d\ar@{--}[dr]	&*={\bullet}\d	\\
*={\bullet}\d\ar@{--}[ur]	&*={\bullet}\d			&*={\bullet}\d	\\
*\txt{$\bullet$}		&*\txt{$\bullet$}		&*\txt{$\bullet$ }			
}
\qquad\qquad\qquad\qquad
\xymatrix@C=0.9pc@R=1pc{
*={\bullet}\d				&*={\bullet}\d				&	 \\
*={\bullet}\d				&*={\bullet}\d				&*={\bullet}\d	\\
*={\bullet}\d				&*={\bullet}\d				&*={\bullet}\d	\\
*\txt{$\bullet$}\ar@{.}[r]		&*\txt{$\bullet$ }\ar@{.}[ur]		&*\txt{$\bullet$}
}
\end{center}
\caption{$\NMA_1$ is isomorphic to the product of the three depicted NM chains (\cite{ABG11}).
The dashed line on the left is the generator, while the dotted line on the right is $[\alpha]_\equiv$.
}\label{fig:nm1exa}
\end{figure}

A straightforward verification shows that for every assignments
$\mu \colon \Form_1 \to [0,1]$, $\mu(\alpha)<1$.
Moreover $[\alpha]_\equiv$ is a join irreducible element of $\NMA_1$.
Indeed, one can check that for every formula $\psi\in \Form_1$ such that
$[\psi]_\equiv\leq[\alpha]_\equiv$, either $[\psi]_\equiv=[\alpha]_\equiv$,
or $[\psi]_\equiv=\bot$. Thus, $\chi(\alpha)= 1$. Compare with Fig.~\ref{fig:nm1exa}.

Since the truth value of $\alpha$ is strictly lower than $1$ under any assignment, but
the Euler characteristic of $\alpha$ is greater than $0$, we can not directly interpret $\chi$ as
a measure of the number of classes of assignments making a formula true. We do not discuss further
the r\^{o}le of Euler characteristic in NM logic here.
Instead, we provide a new valuation that, as we will see later in this section, can be
interpreted similarly to how the Euler characteristic has been interpreted in \god{} logic in \cite{cdm_jmvlsc}.

Let us introduce such a valuation, slightly different from the Euler characteristic,
defined as follows.

\begin{definition}\label{d:euler+}
We define the \emph{idempotent Euler characteristic} $\chi^{+}:\NMA_n \to \R$ as the
valuation on $\NMA_n$ such that
\begin{enumerate}
\item $\chi^{+}(\bot)=0$;
%
\item for each join irreducible element $g \in \NMA_n$,
\begin{align*}
\chi^{+}(g) &=  \begin{cases}
		1 & \text{if } g \odot g = g\,,\\
		0 & \text{otherwise.}
	\end{cases}
\end{align*}
\end{enumerate}
\end{definition}

\begin{remark}
Observe that, if $g$ is a join irreducible element,
but $g \odot g \neq g$, then $g \odot g = \bot$.
\end{remark}
The following proposition highlights a fundamental property of this newly defined valuation.
The name given to the valuation is due to such property.

\begin{proposition}\label{prop:idvalu}
Fix $n\geq 1$. The idempotent Euler characteristic satisfies, for every $x \in \NMA_n$,
\[
\chi^+(x\odot x)=\chi^+(x)
\]
\end{proposition}
\begin{proof}
Let $x\in\NMA_n$. Three cases are to be considered.
\begin{itemize}
\item[1)] $x\odot x = x$.
\item[2)] $x\odot x = \bot$.
\item[3)] $x\odot x = y$, with $y\in\NMA_n$, $y \neq x$, and $y \neq \bot$.
\end{itemize}

If 1) holds the proposition immediately follows. Suppose 2) holds. We need to prove that $\chi^+(x)=\chi^+(\bot)=0$.
First, observe that for every $y\in \NMA_n$ such that $y\leq x$, we have $y\odot y \leq x\odot x$. Thus,
$y\odot y =\bot$. Let $G=\{g_1,\dots,g_m\}$ be the poset of join irreducibles of $\NMA_n$ such that $g_i\leq x$.
Note that $x=\bigvee_{i=1}^m g_i$. We proceed by induction on the structure of $G$. If $m=1$, then $x$ is a join irreducible (an atom
of $\NMA_n$),
$G=\{x\}$, and $\chi^+(x)=0$. Let $m\geq 2$, and suppose (inductive hypothesis) that the proposition holds
for every element $y=\bigvee_{g\in G'} g$, with $G'\subsetneq G$. Suppose $x$ is not a join irreducible (otherwise,
the result follows by Definition \ref{d:euler+}). Say, without loss of generality, that $g_m$ is maximal in $G$, and let $y=\bigvee_{i=1}^{m-1} g_i$.
By Equation \eqref{eq:valuation},
\[
\chi^+(x)=\chi^+(g_m)+\chi^+(y)-\chi^+(g_m \wedge y)
\]
By Definition \ref{d:euler+}, $\chi^+(g_m)=0$. Further, by inductive hypothesis,
$\chi^+(y)=0$. Let $G'$ be the poset of join irreducible $g$ of $\NMA_n$ such that $g\leq y$. Since $g_m$ is join irreducible, and it is maximal in $G$, $y\lneq g_m$. Thus, $G'\subsetneq G$. By inductive hypothesis,
$\chi^+(g_m \wedge y)=0$. We conclude $\chi^+(x)=0$.

Suppose, finally, that 3) holds. Let $z=\neg x \wedge x$. By monotonicity of $\odot$, we obtain $z\odot z = \bot$.
Thus, $\chi^+(z)=0$. Moreover, $y\wedge z \leq z$, thus $(y\wedge z) \odot (y\wedge z)=\bot$. Therefore,
$\chi^+(y\wedge z)=0$.  Using the subdirect representation, one can see that $x=y\vee z$. We obtain
\[
\chi^+(x)=\chi^+(y)+\chi^+(z)-\chi^+(y\wedge z)=\chi^+(y)\,,
\]
and the proposition is proved.
\end{proof}


We do not provide here an example of the values of the idempotent Euler characteristic
on a free NM algebra, because of the dimension of such structures ($\NMA_1$ has 48 elements).
However, a clarifying example is depicted in Fig. \ref{fig:nmmeno1chi}, for the case of $\NMA^-_1$.

\begin{lemma}\label{l:nm}Fix integer $n \geq 1$, and let $x\in \NMA_n$.
Then, $\chi^{+}(x)$ equals the number of minimal idempotent join-irreducible elements
$g\in \NMA_n$ such that $g\leq x$.
\end{lemma}
\begin{proof}
Let $x\in \NMA_n$. If $x=\bot$ the Lemma trivially holds.
Suppose $x \odot x = \bot$, with $x\neq \bot$.
By Proposition \ref{prop:idvalu}, $\chi^+(x)=0$.
Observe that for all $y\leq x$, $y\odot y \leq x \odot x$,
and thus $y\odot y = \bot$. That is, no idempotent element, except $\bot$, is under $x$,
as desired.

Suppose now $x \odot x \neq \bot$. Let $F$ be the forest of all idempotent join
irreducible elements $g\in \NMA_n$ such that $g\leq x$.
Since $x \odot x \neq \bot$, we have $F\neq \emptyset$.
Recall that $x=\bigvee_{g\in F} g$. We proceed by induction on the structure
of $F$. If $F$ has only one element, then $F=\{x\}$. By Definition \ref{d:euler+},
$\chi^+(x)=1$, as desired.

Let now $|F|>1$, let $l\in F$ be a maximal element of $F$, let $F^-=F\setminus \{l\}$,
and let $x^-$ be the join of the elements of $F^-$. Observe that $x=x^- \vee l$.
Denote by $M$ and $M^-$ the number of minimal elements of $F$, and $F^-$, respectively.

If $l$ is a minimal element of $F$, then $M=M^- + 1$. Let $l^-=l\wedge x^-$.
One can check (for instance, using the subdirect representation), the $l^-$ satisfies
$l^- \odot l^-=\bot$. Thus, by Proposition \ref{prop:idvalu}, $\chi^{+}(l^-)=0$.
By \eqref{eq:valuation}, using the inductive hypothesis, we have
$\chi^{+}(x)=\chi^{+}(l)+\chi^{+}(x^-)-\chi^{+}(l^-)=1 + M^- - 0=M$, as desired.

If $l$ is not a minimal element of $F$, then $M=M^-$. Let $l^- =l\wedge x^-$.
Clearly, the forest of idempotent join irreducible elements under $l$ forms a chain, we
denote $L$.
Moreover, one easily see that the forest of idempotent join irreducible elements under $l^-$
is the chain $L\setminus \{l\}$. Thus, $\chi^{+}(l^-)=1$.
By \eqref{eq:valuation}, we have
$\chi^{+}(x)=\chi^{+}(l)+\chi^{+}(x^-)-\chi^{+}(l^-)=1 + M^- - 1=M$, as desired.
\end{proof}

\begin{lemma}\label{lem:bijection}
Fix $n\geq 1$, and let $\varphi\in \Form_n$. Let $O(\varphi,n)$ be the set of assignments
$\mu:\Form_n \to \{0,\frac{1}{2},1\}$ such that $\mu(\varphi)=1$.
Then, there is a bijection between $O(\varphi,n)$
and the set of minimal idempotent join irreducible elements $g\in \NMA_n$ such that
$g \leq [\varphi]_\equiv$.
\end{lemma}

\begin{proof}
Equipping $\{0,\frac{1}{2},1\}$ with the structure of an NM algebra, the resulting
chain will be isomorphic to the three-element NM algebra $\mathbf{3}$.

Fix an assignment $\mu:\Form_n\to \{0, \frac{1}{2}, 1\}$. Then,
there exists a unique homomorphism $h_\mu: \NMA_n\to\mathbf{3}$ defined by
\begin{align}\label{eq:bi1}
h_\mu([\varphi]_\equiv)= \mu(\varphi).
\end{align}

Conversely, for every $h:\NMA_n \to\mathbf{3}$ we can define a unique
assignment $\mu_h:\Form_n\to \{0, \frac{1}{2}, 1\}$ such that
\begin{align}\label{eq:bi2}
\mu_h(\varphi) = h([\varphi]_\equiv).
\end{align}

This yields a bijection between assignments $\mu:\Form_n \to \{0,\frac{1}{2},1\}$
and NM homomorphisms $h:\NMA_n \to \mathbf{3}$.
In particular, consider that $\mu_h(\varphi)=1$ if and only if $h_\mu([\varphi]_\equiv)=1$.
Moreover, $h_\mu^{-1}(1)$ is a prime filter $\mathbf{p}_{h_\mu}$ in $\NMA_n$.

By Lemma~\ref{lem:maxquotient} and the fact that $h_\mu$ is an NM algebra homomorphism,
$\mathbf{p}_{h_\mu}$ has to be maximal.
Hence, for every $\mu \in O(\varphi,n)$ we can associate the minimal idempotent join irreducible element
in $\NMA_n$ that generates $\mathbf{p}_{h_\mu}$.

Conversely, for every $\mathbf{p}$ maximal prime filter in $\NMA_n$
there exists an NM algebras homomorphism
$ h_{\mathbf{p}}:\NMA_n\to\mathbf{3},$
induced by the natural quotient map $\NMA_n\to\NMA_n/\mathbf{p}$
composed with the embedding $\NMA_n/\mathbf{p} \to\mathbf{3}$ given by Lemma~\ref{lem:maxquotient}.
Thanks to the bijection established by \eqref{eq:bi1} and \eqref{eq:bi2},
we are able to associate an assignment $\mu_{h_{\mathbf{p}}}$
with every minimal idempotent join irreducible element $p$ in $\NMA_n$.
And the Lemma is settled.
\end{proof}

Combining Lemma \ref{l:nm} and Lemma \ref{lem:bijection}
we can now state our main result.

\begin{theorem}\label{t:nm}Fix an integer $n \geq 1$.
For any formula $\varphi \in \Form_n$, the valuation
$\chi^{+}(\varphi)$ equals the number of assignments $\mu
\colon \Form_n \to \{0,\frac{1}{2},1\}$ such that $\mu(\varphi)=1$.
\end{theorem}

\begin{remark}
If $\varphi$ is a tautology in NM logic, then
$\chi^+(\varphi)=3^n$.
\end{remark}

\section{Valuations in NM\textsuperscript{-} logic}\label{s:NM-}

As mentioned in Section \ref{sec:NM},
NM$^{-}$ is the schematic extension of NM logic obtained adding the axiom
$\neg(\neg x^2)^2 \leftrightarrow (\neg(\neg x)^2)^2$.
On the algebraic side we have that an NM algebra is an NM$^{-}$ algebra
if and only if it does not have a negation fixpoint.
Since Definitions \ref{d:val_on_NM algebra} and \ref{d:euler+} easily apply
 to the NM$^{-}$ case, we can consider the idempotent Euler characteristic
on free $n$-generated NM$^{-}$ algebras.
As we will see later in this Section, the results we obtain in this case are interesting,
although easy corollaries of the results obtained in the previous Section.

First of all, observe that Proposition \ref{prop:idvalu} and Lemma \ref{l:nm}
clearly hold on $\NMA^-_n$ algebras. Furthermore, we can easily adapt
Lemma~\ref{lem:maxquotient} (and its proof) to $\NMA^-_n$ algebras, as follows.

\begin{lemma}\label{lem:maxquotientNM-}
Let $\mathbf{2}$ be the two-elements NM chain.
Then, given a finite NM$^{-}$ algebra $\mathbf{A}$ and a maximal prime filter $\mathbf{p}$,
the quotient $\mathbf{A}/\mathbf{p}$ is isomorphic to $\mathbf{2}$.
\end{lemma}

Appealing at the proof of Lemma~\ref{lem:bijection},
given a maximal prime filter $\mathbf{p}$, there exists an embedding
from the quotient $\NMA_n^{-}/\mathbf{p}$ to the two-elements NM chain $\mathbf{2}$.
Lemma~\ref{lem:bijection} thus takes the following form, in the NM$^-$ case.

\begin{lemma}\label{lem:bijection_NM-}
Fix $n\geq 1$, and let $\varphi\in \Form_n$. Let $O(\varphi,n)$ be the set of assignments
$\mu:\Form_n \to \{0,1\}$ such that $\mu(\varphi)=1$.
Then, there is a bijection between $O(\varphi,n)$
and the set of minimal idempotent join irreducible elements $g\in \NMA^-_n$ such that
$g \leq [\varphi]_\equiv$.
\end{lemma}

This fact, together with a revised version of Lemma~\ref{l:nm}, allow us to restate our main theorem
for NM$^-$ logic.


\begin{theorem}\label{t:nm_meno}Fix an integer $n \geq 1$.
For any formula $\varphi
\in \Form_n$, the valuation
$\chi^{+}(\varphi)$ equals the number of assignments $\mu
\colon \Form_n \to \{0,1\}$ such that $\mu(\varphi)=1$.
\end{theorem}

\begin{remark}
If $\varphi$ is a tautology in NM$^{-}$, then $\chi^{+}(\varphi)=2^n$.
\end{remark}


\begin{example}
Consider the subdirect representation of $\NMA_1$ given in Fig.~\ref{fig:nm1exa}.
Since the free $1$-generated NM$^{-}$ algebra is a subalgebra of $\NMA_1$,
we can obtain $\NMA_1^{-}$ by removing the three elements NM chain
(it is the only NM chain in the subdirect product with a negation fixpoint).
Indeed, $\NMA_1^{-}$ is obtained as a product of the two $1$-generated four-elements NM chains.
In Fig.~\ref{fig:nmmeno1chi} the order structure of $\NMA_1^{-}$
has been labelled with the values given by the idempotent Euler characteristic.


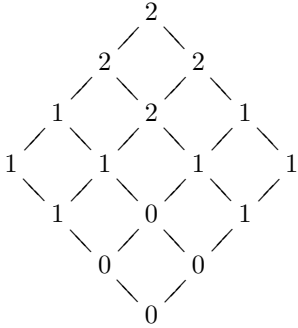
\begin{figure}[h!]
\begin{center}\leavevmode
\xymatrix@C=0.6pc@R=0.6pc{
	&	&			&2\dl\dr	&			&	&	\\
	&	&2\dl\dr	&			&2\dl\dr	&	&	\\
	&1\dl\dr&		&2\dl\dr	&			&1\dl\dr&\\
1\dr&	&1\dl\dr	&			&1\dl\dr	&	&1\dl	\\
	&1\dr&		&0\dl\dr	&			&1\dl	&	\\
	&	&0\dr		&		&0\dl		&	&	\\
	&	&			&0		&			&	&	
}
\end{center}
\caption{The order structure of $\NMA_1^{-}$. Elements are labelled with their
idempotent Euler characteristic.}\label{fig:nmmeno1chi}
\end{figure}

\end{example}

\section{Conclusion, and further work}
Our brief discussion on the (classical) Euler characteristic lead to the conclusion
that a proper logical meaning for such valuations does not follow the intuition
of \cite{cdm_jmvlsc}. We think a deeper investigation deserve to be done.

Further research also has to be done in order to obtain more expressive valuations,
generalizing the idempotent Euler characteristic. Indeed,
as in the G\"{o}del logic case, the study of $k$-valued extensions of NM logic seems to be
a feasible task.

Finally, an approach similar to the one presented here
can be applied to other logics lying in the same hierarchy of G\"odel and NM logics.
An example is NMG logic \cite{WWP05}, the logic of the ordinal sum of G\"odel and NM standard chains.
The study of the Euler characteristic, or some modified versions of such valuation, on NMG algebras
is a natural prosecution of this work.
In order to address the more difficult case given by WNM logic,
a useful and clarifying intermediate step is the study of RDP logic \cite{W07}.
Indeed, the structure of join irreducible elements of RDP logic has already been investigated in
\cite{BV11}, while a poset representations
of its free $n$-generated algebras has been provided in \cite{V10}.

\section*{Acknowledgment}
We thank Stefano Aguzzoli and Vincenzo Marra for many useful discussions on the topics of this work.

The authors were supported by the MIUR-FIRB research project PNCE - Probability theory of non-classical events.
Valota acknowledges also partial support from a Marie Curie INdAM-COFUND Outgoing Fellowship. The research reported in this paper was carried out while Valota was a postdoc fellow of the Dipartimento di Scienze Teoriche e Applicate (Universit\`{a} dell'Insubria), supported by the  PNCE research project.

%
%
%
%

\bibliographystyle{IEEEtranS}
\bibliography{cv_ISMVL2015}
%
%
%

\end{document}